\documentclass[]{jfm}

\usepackage{mathtools}

\usepackage{graphicx}
\usepackage{newtxtext}
\usepackage{newtxmath}
\usepackage{natbib}
\usepackage{hyperref}
\hypersetup{
    linkcolor  = black,
    colorlinks = true,
    urlcolor   = black,
    citecolor  = black,
}
\newcommand{\RomanNumeralCaps}[1]
\linenumbers

\title{Emergent vorticity asymmetry of one and two-layer shallow water system captured by a next-order balanced model}

\author{Ryan Sh\`iji\'e D\`u \aff{1}
  \corresp{\email{ryan\_sjdu@nyu.edu}}, K. Shafer Smith\aff{1}}

\affiliation{\aff{1}Center for Atmosphere Ocean Science, Courant Institute of Mathematical Sciences, New York University, New York, NY 10012, USA}

\newcommand{\de}{\mathrm{d}}
\newcommand{\DD}{\mathrm{D}}
\newcommand{\pe}{\partial}

\newcommand{\ve}[1]{\boldsymbol{#1}}
\newcommand{\qdt}[1]{\quad \mbox{#1} \quad}
\newcommand{\mean}[1]{\left\langle {#1} \right\rangle}

\newcommand{\pl}{\textsuperscript{+1}}
\newcommand{\Ro}{\{\varepsilon\}}
\newcommand{\Ub}{\left\{\textit{Bu}^{-1}\right\}}
\newcommand{\Bu}{\left\{\textit{Bu}\right\}}
\newcommand{\Bt}{\left\{\gamma\varepsilon\right\}}
\newcommand{\Gm}{\left\{\gamma\right\}}
\newcommand{\mcal}{\mathcal}

\begin{document}
\maketitle

\begin{abstract}
{The turbulent evolution of the shallow water system exhibits asymmetry in vorticity. This emergent phenomenon can be classified as ``balanced'', that is, it is not due to the inertial-gravity wave modes. The Quasi-Geostrophic (QG) system, the canonical model for balanced motion, has a symmetric evolution of vorticity, thus misses this phenomenon. Here we present a next-order-in-Rossby extension of QG, QG\pl, in the shallow water context. We recapitulate the derivation of the model in one-layer shallow water grounded in physical principles and provide a new formulation using ``potentials''. Then, the multi-layer extension of the SWQG\pl~model is formulated for the first time. The SWQG\pl~system is still balanced in the sense that there is only one prognostic variable, potential vorticity (PV), and all other variables are diagnosed from PV. It filters out inertial gravity waves by design. This feature is attractive for modeling the dynamics of balanced motions that dominate transport in geophysical systems. The diagnostic relations connect ageostrophic physical variables and extend the massively useful geostrophic balance. Simulations of these systems in classical set-ups provide evidence that SWQG\pl~captures the vorticity asymmetry in the shallow water system.  Simulations of freely decaying turbulence in one-layer show that SWQG\pl~can capture the negatively skewed vorticity, and simulations of the nonlinear evolution of a baroclinically unstable jet show that it can capture vorticity asymmetry and finite divergence of strain-driven fronts. 
}
\end{abstract}



\section{Introduction}
The shallow water system has been a crucial workhorse model for our understanding of geophysical fluid dynamics (GFD) \citep[e.g.][]{Zeitlin_18}. It simplifies the more complex GFD equations (e.g., the Boussinesq system) by assuming that one can represent the vertical variation using only one (or a few) stacked stratified layers. This simplification allows for easier numerical and theoretical investigations of its properties, and in many cases shallow water captures the realistic properties of geophysical phenomena in the atmosphere and ocean of Earth and other planets \citep[e.g.][for just a few]{DowlingIngersoll_89a,PolvaniEtAl_94,ChoPolvani_96,LambaertsEtAl_12,BembenekEtAl_20}. 

A further simplification of the shallow water system is the Shallow Water Quasi-Geostrophic equation (SWQG) \citep{Vallis_17,Zeitlin_18}. It assumes rotation dominates and thus the geostrophic balance approximation is valid and the Rossby number is small. SWQG is then an asymptotic approximation of the shallow water model in the limit of vanishing Rossby number. It is a prototypical example of a balanced model, in that it can be written with only a single prognostic variable, potential vorticity (PV), from which all other physical variables can be diagnosed. This has the consequence that {inertial-}gravity waves are not permitted in its dynamics. SWQG (and its special case, the two-dimensional Euler {equations}) {can capture the dynamics of shallow water in surprisingly wide parameter regime \citep{SugimotoEtAl_07}, and} it has rich turbulence phenomena that closely resemble the balanced dynamics of rotationally dominated atmospheres and oceans \citep[e.g.][]{GillEtAl_74,Salmon_80,MaltrudVallis_91,SmithEtAl_02,DritschelMcIntyre_08,EarlyEtAl_11}. The study of SWQG has led to progress in many practical geophysical applications, including the parametrization of eddies \citep{HeldLarichev_96,ThompsonYoung_07,SrinivasanYoung_14,GalletFerrari_20,GalletFerrari_21}, the interpolation and interpretation of satellite altimetry measurement of sea surface height \citep{Stammer_97,CheltonEtAl_07,UbelmannEtAl_15}, the analysis of ocean coherent vortices over topography features \citep{BrethertonHaidvogel_76,CarnevaleFrederiksen_87,SiegelmanYoung_23,LaCasceEtAl_24}, the statistics of atmospheric blocking \citep{MarshallMolteni_93,LucariniGritsun_20}, and the mechanics of jets and spots on gas giants \citep{TurkingtonEtAl_01,MajdaWang_06,SiegelmanEtAl_22,PizzoSalmon_24}, to name but a few.

Despite the range of phenomenology SWQG captures, it still misses many features of the full shallow water model, even those that might be categorized as ``balanced''. SWQG misses the asymmetry between cyclonic and anticyclonic vortices. \citet{PolvaniEtAl_94} note that the vorticity skews negative in a freely decaying simulation of the shallow water system. \citet{AraiYamagata_94,GravesEtAl_06} further analyze the detailed free evolution of vortices in the shallow water model and conclude that anticyclones are more robust to small-scale perturbation and to the influence of other vortices, providing a plausible mechanics of the negative skewness of vorticity. 
For an ocean gyre forced by wind, \citet{ThiryEtAl_24} found the shallow water model version exhibits asymmetric drift of the strong eastward jet, while the SWQG version lacks this asymmetry. 
For multi-layer shallow water, \citet{LambaertsEtAl_12} find that the evolution of baroclinically unstable jets in two-layer shallow water exhibits complex vorticity asymmetry evolution, further complicated by their inclusion of moisture.

In this paper, we aim at a balanced model of the shallow water system that could capture the evolution of asymmetry between cyclonic and anticyclonic vortices. {We focus on the parameter space where SWQG is formally valid but lacking in captured phenomena.} As a by-product, we also wish to capture the finite divergence of shallow water flows, missed by SWQG. 
Some balanced models already exist in the literature. 
One line of models includes the frontal geostrophic model by \citet{Cushman-Roisin_86} and the generalized geostrophic equation by \citet{Cushman-RoisinTang_89,Cushman-RoisinTang_90}. These models capture a more extended parameter space of the shallow water model by allowing for large height deviation and can capture vorticity asymmetry. However, they are not higher order in Rossby number than SWQG, and they all insist that the velocities are in geostrophic balance with the height field and thus do not allow for divergent flow. 
Another model is the Balance Equation \citep{AllenEtAl_90,SpallMcwilliams_92}. It is a capable model of shallow water in the QG regime that includes higher-order effects and is useful as a way to initialize a shallow water simulation with minimal generation of gravity waves. We use it to generate the initial conditions for our freely decaying simulation. However, evolving the Balance Equation is challenging. The system has two time derivatives but is still a balanced model in the sense that there is only one prognostic variable. This is because the two time evolution equations are not independent from each other but coupled in complex nonlinear ways. The time evolution usually needs iterative solvers per timestep \citep{BarthEtAl_90}. This poses challenges when using it to study the evolution of the shallow water model {and fully turbulent simulations of it are rare}. 
{Yet another model is} the semi-geostrophic model of \citet{Hoskins_75} adapted to the shallow water context \citep{ClokeCullen_94,RoulstoneSewell_96}. It has features and deficiencies similar to those of the Boussinesq semi-geostrophic model. It has energy and PV conservation and can capture vorticity asymmetry. However, it is not of higher order in Rossby number. The semi-geostrophic model is less accurate in modeling curved fronts. The kinetic energy of the semi-geostrophic model includes only the geostrophically balanced component, making it hard to interpret. The inversion from PV involves a change of coordinate and solving a nonlinear elliptic Monge-Amp\`ere problem, both are hard to do numerically. {Practically, \cite{Fletcher_04} explored its utility as a model for data assimilation in numerical weather prediction. Theoretically,} the mathematical challenge of analyzing a Monge-Amp\`ere problem coupled to a transport equation has inspired deep mathematical analysis of the semi-geostrophic model for shallow water \citep[see][and references therein]{CullenGangbo_01,DePhilippisFigalli_14}. 
{Finally, the class of optimal balance algorithms deals more broadly with the issue of optimal separation of balanced and unbalanced components of a \emph{snapshot} of the flow, going beyond asymptotics \citep{ViudezDritschel_04,MasurOliver_20,ChoukseyEtAl_23}. Their focus is not on writing a prognostic system that is balanced for all time.}

{We explore and expand on a balanced model, SWQG\pl, that extends the SWQG into one higher order in Rossby number {in an asymptotically consistent manner}. \citet{WarnEtAl_95} first derived SWQG\pl~for the one-layer shallow water system, and \citet[][hereafter V96]{Vallis_96a} extended it to include the $\beta$-effect.} The higher-order-in-Rossby terms model the asymmetric correlation between height deviation and vorticity and the cyclostrophic balance. 
Recently \citet{ChoukseyEtAl_23} has studied a variant of the \citet{WarnEtAl_95} model's ability for balanced initialization of the shallow water model {and has {compared} it to the optimal balance algorithm of \citet{MasurOliver_20}. } 
{In this paper, we rewrite the \citet{WarnEtAl_95} model using ``potentials'', inspired by \citet{MurakiEtAl_99}'s version of QG\pl~of the Boussinesq system.} 
This new potential form allows us to extend the model to multi-layer for the first time. {We show through nonlinear simulation of the one and two-layer SWQG\pl~system that the model captures the asymmetric evolution of vorticity in the one and two-layer shallow model, as well as flow fields with finite divergence. }

SWQG\pl, like QG\pl~in the Boussinesq context, is a PV-based balanced model, where the prognostic variable is PV and all other variables are diagnosed from it. 
\citet{WarnEtAl_95} first considered abstractly the problem of constructing higher-order balanced models and determined that a stable model should not expand the prognostic variable. For SWQG\pl~this means PV is not expanded in the small Rossby number. All other variables are expanded and ``slaved'' to PV by the principle of ``PV inversion'' of \citet{HoskinsEtAl_85}.
Choosing PV as the central variable is also based on the fact that PV has been crucial in understanding many features of the shallow water model and SWQG \citep{BrethertonHaidvogel_76,HaynesMcIntyre_87,Hoskins_91,DritschelMcIntyre_08}. 
{In addition to theoretical appeal, having PV as the only prognostic variable guarantees that the model is free of inertial-gravity waves, just like SWQG. The additional diagnostic relations of ageostrophic but balanced velocities can be useful in their own right, allowing for inference of ageostrophic dynamics from sparse observations. In particular, this can be useful in the new era of high-resolution satellite altimetry observation where geostrophic balance fails in the submesoscale \citep{PenvenEtAl_14}.
}

{The rest of the paper is organized as follows. In Section \ref{sec:single}, we rederive the one-layer SWQG\pl~model using the potentials, which is an alternative form of the model of \citet{WarnEtAl_95} and V96.} We explore its turbulent free decay, compared to the corresponding shallow water model. It is shown that SWQG\pl~captures the emergent negative vorticity skewness in the shallow water model. We also track the energy evolution. It is shown that while theoretical energy conservation law is lacking, the energy of SWQG\pl~models the energy evolution of the full shallow water model well. 
In Section \ref{sec:multi}, we extend the SWQG\pl~model to many layers. A simulation of a baroclinically unstable jet demonstrates SWQG\pl~can capture the vorticity asymmetry in two layers. Contrary to the results of the one-layer simulations, the two-layer simulation shows the initial baroclinic growth stage has a cyclonic bias for vorticity. It also shows patterns of vorticity and divergence similar to strain-driven fronts. 
The vortex stretching mechanism leads to the cyclonic vorticity bias. 
We conclude in Section \ref{sec:conclude} and discuss possible applications and extensions to the model.

\section{Single-layer shallow water}\label{sec:single}
This section re-derives the SWQG\pl~model, providing an alternative to the \citet{WarnEtAl_95}'s approach. By using ``potentials'' inspired by the \citet{MurakiEtAl_99} model for Boussinesq, our form of SWQG\pl~involved three elliptic inversions where the elliptic operator to be inverted is the same as the one in SWQG, the Screened Poisson operator. We simulate the free decay of random balanced initial conditions of SWQG\pl~model as well as the shallow water model and compare their turbulent statistics. SWQG\pl~can capture the vorticity asymmetry in the shallow water model.

\subsection{Derivation of the SWQG\pl~model}
\subsubsection{The shallow water system and its properties}
We start with the one-layer shallow water model {on a $\beta$-plane and mean layer depth of $H$}. {The equations for horizontal velocities} $(u,v)$ and water layer height perturbation $h$ are
\begin{subequations}\label{eq:oneSW}
\begin{align}
    & \Ro\left( \frac{\DD u}{\DD t} \right) - (f+\Bt \beta y)v = -g h_x\\
    & \Ro\left( \frac{\DD v}{\DD t} \right) + (f+\Bt \beta y)u = -g h_y\\
    & \Ro\left(\frac{\DD h}{\DD t}+h\nabla\cdot u\right) + \Bu H\nabla\cdot \ve u = 0.\label{eq:oneSW_h}
\end{align}
\end{subequations}
To facilitate future asymptotic analysis but also keep the physical constants for interpretability, we write the equation using the notation of V96. All nondimensional constants are enveloped in curly brackets, and one can ignore them if one wants only the physical equations. One can instead obtain the nondimensional equations by removing all physical constants, here $f$, $\beta$, $g$, and $H$. The nondimensional numbers in this model are the Rossby number:
\begin{align}
    \varepsilon := \frac{U}{fL},\label{eq:Ro_def}
\end{align}
the Burger number:
\begin{align}
     \textit{Bu} := \frac{gH}{f^2L^2},\label{eq:Bu_def}
\end{align}
and the 
Charney--Green number
\begin{align}
    \gamma := \frac{\beta L^2}{U}
\end{align}
which capture the effects of $\beta$. In this paper, $\beta$ is retained in the derivation, but its effect is not explored in the simulations. 
Here we scale $h$ using geostrophic balance
\begin{align}
    h \sim \frac{fUL}{g} \qdt{and} \frac{h}{H} \sim \frac{\varepsilon}{\textit{Bu}}.\label{eq:eta_scaling}
\end{align}

The shallow water system conserves the potential vorticity (PV)
\begin{align}
    &\frac{\DD Q}{\DD t} = 0 \qdt{with} Q = \frac{f+\Ro\zeta + \Bt \beta y}{H+\{\varepsilon/\textit{Bu}\}h}.\label{eq:SWPV_full}
\end{align}
As a consequence, it conserves the total potential enstrophy density:
\begin{align}
        \text{Potential Enstrophy} = \frac{1}{2}\mean{ (H+\{\varepsilon/\textit{Bu}\}h) Q^2}\label{eq:poten_ens}
\end{align}
where $\mean{\cdot}$ is the area average:
\begin{align}
    \mean{\cdot} = \frac{1}{A} \iint_A \cdot \;\de x\de y.
\end{align}
It also conserves the total energy made up of eddy kinetic energy (EKE) and available potential energy (APE) density:
\begin{subequations}\label{eq:SW_energies}
\begin{align}
    \text{Total energy} &= \text{EKE}+\text{APE}\\
    \qdt{where}\text{EKE} &= \frac{1}{2}\mean{(H+\Ro h)(u^2+v^2)}\\
    \text{APE} &= \Ub\frac{1}{2}\mean{gh^2}.
\end{align}
\end{subequations}

To prepare for the derivation of the SWQG\pl~model, we propose a form of representing the three variables of shallow water using three ``potentials''
\begin{subequations}
\begin{align}
    u &= -\Phi_y-F,\\
    v &= \Phi_x-G,\\
    h &= \frac{f}{g}\Phi-\Bu\frac{H}{f}G_x+\Bu\frac{H}{f}F_y,
\end{align}
\end{subequations}
where we assume the scaling
\begin{align}
    \Phi\sim UL \qdt{and} G = F \sim U.
\end{align}
This procedure and naming are inspired by QG\pl~for the Boussinesq model where the three potentials are the vector potential form of the three-dimensional incompressible velocity field \citep{MurakiEtAl_99,DuEtAl_24}. Here, for the shallow water case, the natural counterpart is the two-dimensional Helmholtz decomposition as adopted by \citet{WarnEtAl_95} and V96. Though our potentials might look unnatural at first glance, it is useful for the derivation of SWQG\pl, where it turns out the elliptic inversion problems for all three potentials are the same as SWQG's Screened Poisson problems. This uniformity guides the derivation of SWQG\pl~in the more complex multi-layer case.

\subsubsection{The shallow water QG and QG\pl~model}\label{sec:sing_QGp1_deri}
SWQG\pl~is a model that makes the same asymptotic assumption as SWQG but extends it to the next level in Rossby number. That is, we assume the small parameter
\begin{align}
    \varepsilon \ll O(1).
\end{align} 
and
\begin{align}
    \textit{Bu} = O(1) \qdt{and} \gamma \leq O(1).
\end{align}
Using the small parameter, we can asymptotically expand the PV:
\begin{subequations}\label{eq:PV_expan}
\begin{align}
    HQ &= \frac{f+\Ro\zeta+\Bt \beta y}{1+\{\varepsilon/\textit{Bu}\}h/H}\\
    &= [f+\Ro\zeta+\Bt \beta y]\left(1-\left\{\frac{\varepsilon}{\textit{Bu}}\right\}\frac{h}{H}+\left\{\frac{\varepsilon^2}{\textit{Bu}^2}\right\}\frac{h^2}{H^2}\right)+O(\Ro^3)\\
    &= f+\Ro\left[(\zeta+\Gm \beta y)-\Ub\frac{fh}{H}\right]\\
    &\qquad+\Ro^2\left[\Ub^2\frac{fh^2}{H^2}-\Ub\frac{(\zeta+\Gm \beta y)h}{H}\right]+O(\Ro^3)\\
    &= f+\Ro q+O(\Ro^3)
\end{align}
\end{subequations}
where we define $q$ in the last line.
We also expand the potential forms under this QG scaling regime:
\begin{subequations}\label{eq:streamexp_onelay}
\begin{align}
    u &= -\Phi^0_y+\Ro\left(-\Phi^1_y-F^1\right),\\
    v &= \Phi^0_x+\Ro\left(\Phi^1_x-G^1\right),\\
    h &= \frac{f}{g}\Phi^0+\Ro\left(\frac{f}{g}\Phi^1-\Bu\frac{H}{f}G^1_x+\Bu\frac{H}{f}F^1_y\right).\label{eq:strm_expan}
\end{align}
\end{subequations}
where the zeroth level is just SWQG where all variables are based on the horizontal streamfunction $\Phi^0$.

The SWQG emerges naturally from the above asymptotic expansions. 
It is a PV-based balanced model where the PV is advected by the zeroth order velocities. That is, we have
\begin{align}
    \frac{\DD^0 q}{\DD t} = 0
\end{align}
where
\begin{align}
    u^0 &= -\Phi^0_y, \qdt{and} v^0 = \Phi^0_x.
\end{align}
To follow the principle of PV inversion, we diagnose $\Phi^0$ from the PV, working with only the zeroth QG level of \eqref{eq:PV_expan}:
\begin{subequations}
\begin{align}
    q &= (\zeta^0+\Gm \beta y)-\Ub\frac{fh^0}{H}\\
    &= \nabla^2\Phi^0-\Ub\frac{f^2}{gH}\Phi^0+\Gm \beta y\label{eq:zeroth_QGinvert}.
\end{align}
\end{subequations}
This is the SWQG equation. It is more common to separate out the $\beta$ term from the QGPV and include it instead in the advection equation. That is, we write
\begin{subequations}
\begin{align}
    &\frac{\DD q}{\DD t} + \Gm\beta v = 0 \label{eq:QG_betainadv}\\
    &\qdt{where} q = \nabla^2\Phi^0-\Ub\frac{f^2}{gH}\Phi^0 = \mcal{S}(\Phi^0).\label{eq:QG_Phi0inv}
\end{align}
\end{subequations}
We define the Screened Poisson operator $\mcal{S}$ for future notational savings. 

To extend the model to the next order in Rossby, we only need to extend the order in the PV inversion. It is worth repeating that following the theoretical argument of \citet{WarnEtAl_95}, PV goes under no asymptotic expansion. The elliptic inversion problem for $\Phi^1$ is obtained from the next order of \eqref{eq:PV_expan}:
\begin{align}
    \mcal{S}(\Phi^1) 
    &= C_q-\Ub^2\frac{f^3}{g^2H^2}(\Phi^0)^2+\Ub\frac{f}{gH}[\nabla^2\Phi^0\Phi^0 + \Gm\beta y\Phi^0].\label{eq:SWpl_Phi1inv}
\end{align}
$C_q$ is the appropriate constant that make $\Phi^1$ to have zero mean 
\begin{align}
    C_q &= \Ub \frac{f}{gH} \mean{\Ub\frac{f^2}{gH}|\Phi^0|^2-\nabla^2\Phi^0\Phi^0 - \Gm\beta y\Phi^0}.\label{eq:Cq_intecond}
\end{align}
$\Phi^0$ and $\Phi^1$ should have zero mean for all time in the simulation to maintain mass conservation \eqref{eq:strm_expan} (assuming {without loss of generality that} $\mean{h}=0$). Therefore for SWQG\pl~the $\Phi^0$ inversion \eqref{eq:zeroth_QGinvert} should use $q-\mean{q}$. Note that with $\beta=0$, the integral of the right-hand side is proportional to the total energy at the QG level after integration by parts
\begin{align}\label{eq:energy_QGlev}
    \text{Total energy}^0 &= \frac{H}{2} \mean{ \Ub \frac{f^2}{gH} |\Phi^0|^2+|\nabla\Phi^0|^2}.
\end{align}

The inversion for $F$ and $G$ is inspired by the derivation of the ``$\omega$-equation'' of shallow water where we form the imbalance equation \citep{HoskinsEtAl_78a}. We first write the $O(\epsilon)$ level of the shallow water system
\begin{subequations}
\begin{align}
    & \frac{\DD^0u^0}{\DD t} - \Gm \beta yv^0 = fv^1 -g h^1_x\\
    & \frac{\DD^0v^0}{\DD t} + \Gm \beta yu^0 = -fu^1-g h^1_y\\
    & \frac{\DD^0h^0}{\DD t} = -\Bu H\nabla\cdot \ve u^1
\end{align}
\end{subequations}
We cancel the time derivative using the thermal wind balance between $v^0$ and $h^0$:
\begin{subequations}
\begin{align}
    &\left(\frac{\DD^0 h^0}{\DD t}\right)_x-\frac{f}{g}\frac{\DD ^0v^0}{\DD t}-\frac{f}{g}\Gm \beta y u^0 \\
    =& -\Bu H(u^1_{xx}+v^1_{xy})-\frac{f}{g}\left(-g h^1_y - fu^1\right).
\end{align}
\end{subequations}
The left-hand side just involved the zeroth order terms represented by $\Phi^0$:
\begin{align}
    &\Ub \frac{1}{H}\left[\left(\frac{\DD^0 h^0}{\DD t}\right)_x-\frac{f}{g}\frac{\DD ^0v^0}{\DD t}-\frac{f}{g}\Gm \beta y u^0\right] = \Ub\frac{f}{gH}\left[J(\Phi_x^0,\Phi^0)+\Gm \beta y \Phi_y^0\right]
\end{align}
and the right-hand side is in fact $\mcal{S}(F^1)$:
\begin{align}
    &-(u^1_{xx}+v^1_{xy})-\Ub\frac{f}{gH}\left(-g h^1_y- fu^1\right) = \mcal{S}(F^1).
\end{align}
Together, we have the elliptic inversion for $F^1$:
\begin{align}
    \mcal{S}(F^1) = \Ub\frac{f}{gH}\left[J(\Phi_x^0,\Phi^0)+\Gm \beta y \Phi_y^0\right].\label{eq:SWpl_F1inv}
\end{align}
A similar derivation gives
\begin{align}
    \mcal{S}(G^1) = \Ub\frac{f}{gH}\left[J(\Phi_y^0,\Phi^0)-\Gm \beta y \Phi_x^0\right].\label{eq:SWpl_G1inv}
\end{align}

In summary, the SWQG\pl~inversions provides a way to obtain the physical variables \eqref{eq:streamexp_onelay} from knowledge of the PV field $q$. It consists of four linear elliptic inversions \eqref{eq:QG_Phi0inv}, \eqref{eq:SWpl_Phi1inv}, \eqref{eq:SWpl_F1inv}, and \eqref{eq:SWpl_G1inv}. These inversions are all the same Screened Poisson problem. This is typical for a model based on asymptotic expansions. The above derivation of SWQG\pl~uses the same ingredient as the model in \citet{WarnEtAl_95} and V96, and they are in fact equivalent. The equivalence of the Boussinesq version of the QG\pl~model of the model in the Boussinesq version of the V96 model is shown in \citet{DuEtAl_24}. We show the equivalency for the shallow water version explicitly in Appendix \ref{app:V96}. {Here, we comment on two points of interest. SWQG\pl~is equivalent to the Bolin–Charney balance equation \citep{Bolin_55,Charney_55a}. 
\begin{align}
    f \zeta^1-g\nabla^2 \eta^1 &= -2J(\Phi^0_x,\Phi^0_y)
\end{align}
where we have set $\beta=0$ for simplicity. This implies SWQG\pl~captures the cyclogeotrophic balance, {which has been useful in diagnosing ageostorphic velocity from sea surface height observations \citep{PenvenEtAl_14}}. Additionally, we can combine the elliptic inversion for $F^1$ and $G^1$ to form an elliptic problem for divergence. 
\begin{align}
    \mcal{S}(u_x+v_y) &= \Ub\frac{f}{gH}\left[J(\Phi^0,\nabla^2\Phi^0)+\Gm \beta \Phi_x^0\right].
\end{align}
This is more commonly called the shallow water $\omega$-equation.}

SWQG\pl~uses PV as the prognostic variable. Therefore, it retains the PV conservation of \eqref{eq:SWPV_full}. However, we have not been able to show that it conserved any definition of energy. {Instead, we will demonstrate numerically that the SWQG\pl~model generally captures the evolution of the shallow water energy \eqref{eq:SW_energies}, compared to the full model. However, it is occasionally not monotonic for the largest Rossby number explored. Since the simulations use dissipation, this does not prove but indicates that SWQG\pl does not conserve the usual shallow water energy.} SWQG\pl~does not conserve potential enstrophy \eqref{eq:poten_ens} either, which additionally requires the height to evolve following the height equation \eqref{eq:oneSW_h}. SWQG\pl's height is inverted from PV and does not necessarily follow \eqref{eq:oneSW_h}. We also show from simulations that the potential enstrophy \eqref{eq:poten_ens} {evolved similarly to a full shallow water simulation}. Overall, even though SWQG\pl~does not explicitly conserve the energy and potential enstrophy, they mirror the behavior of the underlying shallow water model {for long-time simulations}. We argue that theoretical thinking based on the conserved quantities can be applied to the SWQG\pl~simulations.

\subsection{Freely decaying simulations of the one-layer model}\label{sec:free_sim}
The SWQG\pl~model was numerically investigated in V96 as a diagnostic tool. That is, given the true PV field of a shallow water simulation, could SWQG\pl~improve the inverted height and velocities, over QG? The conclusion in V96 is only in certain cases. Here we explore the modeling power of SWQG\pl~to capture the emergent vorticity asymmetry in shallow water. We evolve an ensemble of shallow water and SWQG\pl~simulations from the same random balanced initial conditions and compare their turbulent statistics. 

\subsubsection{Set-ups of the simulations}
The simulations are set in an idealized mid-latitude ocean mixed-layer modeled as a one-and-a-half-layer shallow water system. It is translated to our dimensional parameters of height $H=100 \text{ m}$, Coriolis parameter $f=10^{-4}\text{ s}^{-1}$, and a reduced gravity constant $g = 1\text{ m/s}^2$ (with the usual prime omitted to match the notation of the rest of the section). These parameters imply a deformation radius $\sqrt{gH}/f$ of 100 km. The domain is doubly-periodic with $Lx=Ly=12\pi\times 100$ km. All the freely decaying simulations have the Burger number \eqref{eq:Bu_def} equal to one, that is, we nondimensionalize with $L=100$ km and the nondimensional domain is $12\pi\times 12\pi$. A typical flow speed at energetic regions of the upper ocean (e.g., the Gulf Stream) is around $U=1.2$ m/s. This gives a Rossby number \eqref{eq:Ro_def} of $\varepsilon=0.12$ while a more quiescent region has $U=0.1$ m/s and $\varepsilon=0.01$. Our simulation will explore this range of realistic Rossby numbers. Of course, the power of nondimensionalization allows our simulation to apply to the atmosphere as well. We do not explore higher Rossby numbers since too high a Rossby number while keeping the Burger number fixed will result in $\{\varepsilon/Bu\}h/H>1$, a drying of part of the domain. This is unrealistic for large-scale ocean and atmosphere applications and cannot be simulated using our pseudospectral code. 

The initial conditions are carefully constructed to benefit the study of vortex asymmetry. They should have zero vorticity skewness initially and be well-balanced to avoid excessive gravity-wave emission. For this, we use the procedure based on the Balance Equations used in \citet[(2.5)]{PolvaniEtAl_94}. The vorticity is set to be isotropic Gaussian random fields (thus no skewness) centered around nondimensional size $4$ and nondimensional wavenumber $1.6$, crudely mimicking energy injected by baroclinic instability. The magnitude is determined so that the QG eddy kinetic energy of the vorticity field is one-half:
\begin{align}
    -\frac{1}{2}\mean{\nabla^{-2}\zeta\cdot \zeta} = \frac{1}{2}.
\end{align}
The fixed-point algorithm described in \citet[]{PolvaniEtAl_94} is then performed to solve for the well-balanced divergence and height fields to match the vorticity fields. 

Vorticity asymmetry emerges from the evolution of the shallow water model. We evolve the simulation pseudospectrally using Dedalus \citep{BurnsEtAl_20} for 500 nondimensional time. The horizontal resolution is $1024\times 1024$ Fourier modes. We dissipate the small-scale for numerical stability using a fourth-order hyper-dissipation {just large enough to absorb small-scale noise, as diagnosed by the vorticity spectrum}
\begin{align}
    \nu \nabla^4 (\cdot) \qdt{with {non-dimensional value}} \nu=5.5\times 10^{-6}. 
\end{align}
on $u,v,h$ for the shallow water simulation and on $q$ for the SWQG\pl~simulation. Time-stepping uses the 3rd-order 4-stage diagonally-implicit+explicit Runge-Kutta (DIRK+ERK) scheme \citep{AscherEtAl_97}. The timestep size is determined to have a CFL number of 0.5. 

\subsubsection{Simulation results}
\begin{figure}
    \centering
    \includegraphics{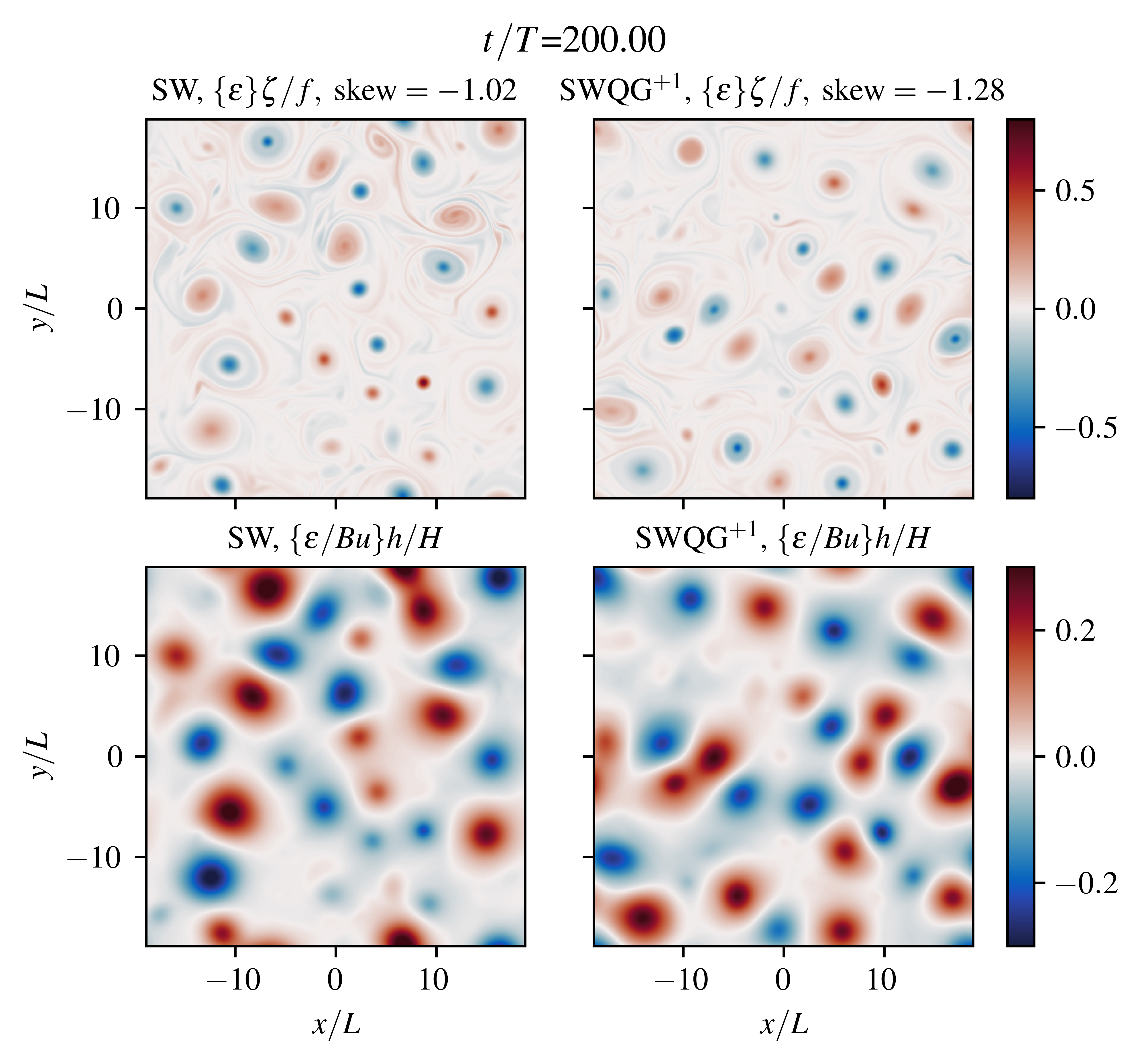}
    \caption{Vorticity ($\Ro\zeta/f$) fields (top) and height ($\{\varepsilon/Bu\}h/H$) fields (bottom) for $\varepsilon=0.1$ at time $t/T=200$ from the shallow water simulation (left) and SWQG\pl (right).}
    \label{fig:snap_t200}
\end{figure}

Figure \ref{fig:snap_t200} shows the comparison of snapshots of vorticity and height for the two models at $t/T=200$ with $\varepsilon=0.1$. The domains are filled with roaming vortices with a bias towards anticyclonic ones. The dominant length scale becomes large compared to the initial condition, which is associated with EKE conversion to APE (cf. Figure \ref{fig:energy_timeline}) and the inverse cascade. {For the one-layer simulations, balanced evolution of vortices does not generate motions with strong divergence, and its effect on the dynamics is minimal. Therefore, we do not further investigate divergence for the one-layer simulations.}

\begin{figure}
    \centering
    \includegraphics{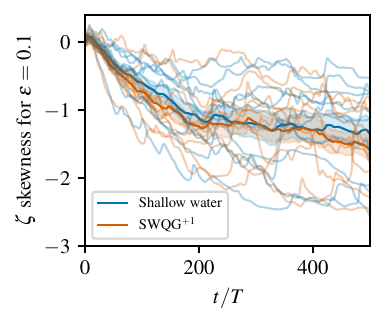}
    \includegraphics{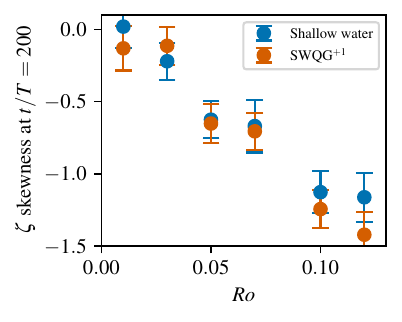}
    \caption{Left: time series of vorticity ($\zeta$) skewness for the $\varepsilon=0.1$ simulations from the shallow water model as well as SWQG\pl. The lighter lines are the individual ensemble members. The darker lines are the ensemble mean, and the $1/\sqrt{10}$ of the ensemble standard deviation is the filled color around the mean.
    Right: the vorticity ($\zeta$) skewness at $t/T=200$ for $\varepsilon=0.01,0.03,0.05,0.07,0.1,0.12$. The error bar is $1/\sqrt{10}$ of the ensemble standard deviation.}
    \label{fig:SW_vortskew_timeline}
\end{figure}

Ensemble statistics are appropriate for studying the emergent vorticity asymmetry of the shallow water model and whether the SWQG\pl~model can capture it. The left of Figure \ref{fig:SW_vortskew_timeline} shows the time evolution of vorticity skewness of the size ten ensemble of simulations of the shallow water model and SWQG\pl, for $\varepsilon=0.1$. The initial conditions are constructed to have symmetric vorticity, and indeed, the vorticity skewness starts at zero for all simulations. 
All simulations' vorticity skewness trends towards negative. The spread between members of the ensembles becomes larger over time, as expected for the chaotic dynamics of shallow water turbulence. However, the ensemble means of vorticity skewness of the two models agree remarkably well during the entire time series. They lie within $1/\sqrt{10}$ times the ensemble standard deviation from each other for most of the time series. We use this range since it is the standard scaling of error of a Monte-Carlo estimate of the ensemble mean. The close agreement indicates strongly that SWQG\pl~model can capture the mechanism that leads to vorticity asymmetry in the shallow water model \citep{AraiYamagata_94,GravesEtAl_06}. 

The agreement extends to all Rossby numbers explored. The right of Figure \ref{fig:SW_vortskew_timeline} shows the ensemble mean and $1/\sqrt{10}$ times the ensemble standard deviation of vorticity skewness for the shallow water model and SWQG\pl~at nondimensional time $t/T=200$. We see that vorticity skewness becomes more negative as the Rossby number gets larger. \citet[FIG. 10]{PolvaniEtAl_94} shows a similar trend except with the Froude number, which is equal to the Rossby number for all our simulations where the Burger number is one. 

\begin{figure}
    \centering
    \includegraphics{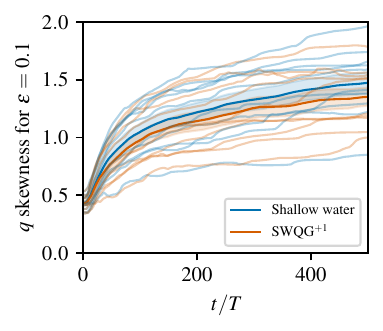}
    \includegraphics{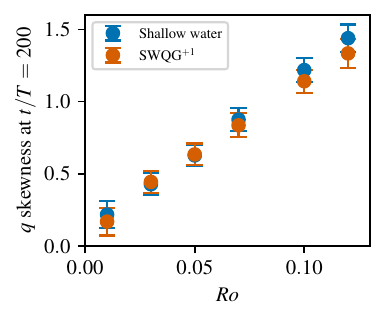}
    \caption{The same as Figure \ref{fig:SW_vortskew_timeline} but for PV ($q$).}
    \label{fig:SW_PVskew_timeline}
\end{figure}

\begin{figure}
    \centering
    \includegraphics{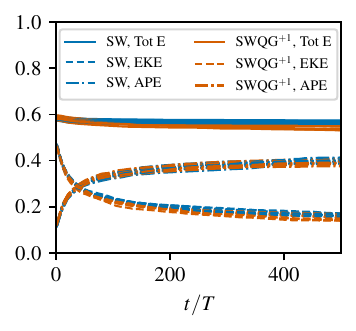}\hspace{3mm}
    \includegraphics{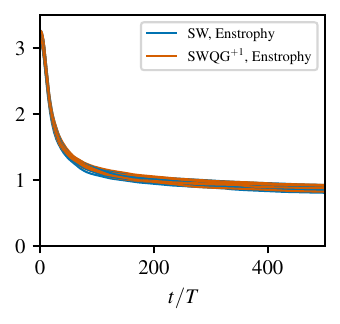}
    \caption{Left: the time series of the total energy, EKE, and APE \eqref{eq:SW_energies} for the $\varepsilon=0.1$ simulations of the shallow water model and SWQG\pl. Right: the time series of the potential enstrophy \eqref{eq:poten_ens}. }
    \label{fig:energy_timeline}
\end{figure}

Since SWQG\pl~is PV-based, we also explore the PV skewness of the evolution. The left of Figure \ref{fig:SW_PVskew_timeline} shows that the evolution of PV skewness again matches well between the shallow water and the SWQG\pl~model, and the right shows the agreement is uniform overall for all Rossby numbers explored. However, the PV skews positive instead. Therefore, the main contributor of the negative vorticity skewness is the nonlinearity in the full shallow-water PV \eqref{eq:SWPV_full}, where positive vortex correlated with negative height deviation and vice versa. In particular the $\zeta$-$h$ correlation term in the asymptotic expansion of $q$ \eqref{eq:PV_expan} scales as
\begin{align}
    \frac{\varepsilon}{Bu} = \frac{U^2}{gH}
\end{align}
which is equal to the Froude number squared. This is consistent with the fact that vorticity skewness is proportional to the Froude number \citep{PolvaniEtAl_94}.

The freely evolving simulation in the shallow water model is a simple setting where we can investigate the properties of the SWQG\pl~model. We have not been able to show that the SWQG\pl~model conserves energy. Instead, we diagnose the energy and show that it {mimics the behavior of the energy of the shallow water model}. The left of Figure \ref{fig:energy_timeline} shows the time series of the total energy as well as the EKE and APE components \eqref{eq:SW_energies} for the evolution of the $\varepsilon=0.1$ ensemble. 
The EKE decreases as it is converted to the APE, a signature of the inverse cascade. The total energy is well-conserved such that more than 90\% of it is in the final state. {The SWQG\pl~total energy evolution closely follows the one for the full shallow water model, and this is true for all Rossby numbers explored (not shown).}
However, it is noticeable that the SWQG\pl~loses more energy, compared to the shallow water model. The shallow water model dissipates the physical variables separately, while the SWQG\pl~model dissipates the PV. Dissipation in the shallow water model does not translate to dissipation of the PV, because of the nonlinearity in the shallow water PV \eqref{eq:SWPV_full}. This is well-known for the Ertel PV of the Boussinesq system, where the dissipation of the physical variables translates to a source term in the PV equations \citep{HaynesMcIntyre_87}. We will not explore the possibility of including source terms in the PV equation in this work any further. The total energy is minimally dissipated, and the dissipation effects are not significant. 

\begin{figure}
    \centering
    \includegraphics{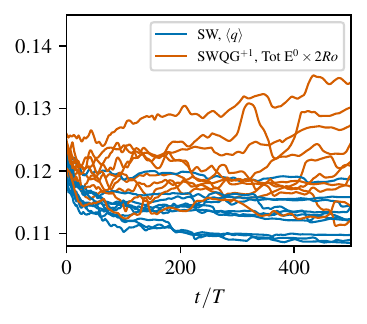}
    \caption{Time series of the total energy at the QG level \eqref{eq:energy_QGlev} in the SWQG\pl~simulations, compared to the mean PV ($\mean{q}$) of the shallow water model.}
    \label{fig:PVmean_timeline}
\end{figure}

It is important to remark that the total energy of the SWQG\pl~model is not monotonic with time, {though this is hard to notice in the timeseries in Figure \ref{fig:energy_timeline}. This implies that the SWQG\pl~model does not conserve the shallow water energy, at least with this choice of dissipation of PV.}
{An alternative reasonable measurement of energy is the total energy at the QG-level \eqref{eq:energy_QGlev}. But it does not even decrease overall and has a much larger ensemble spread, as shown by the time series in Figure \ref{fig:PVmean_timeline}. This further shows our simulated regime is beyond simple QG dynamics.}

Potential enstrophy is effectively dwindled away by the small-scale dissipation, as shown by the right of Figure \ref{fig:energy_timeline}. The evolutions match well between the shallow water model and SWQG\pl. 

We have commented in Section \ref{sec:sing_QGp1_deri} that {for SWQG\pl~}total energy at the QG level \eqref{eq:energy_QGlev} is proportional to $C_q$ \eqref{eq:Cq_intecond}. It is the SWQG\pl~representation of the mean PV ($\mean{q}$) of the shallow water model. 
Figure \ref{fig:PVmean_timeline} compares the two. Overall, the two values are similar. This confirms the assumption made in the derivation that the difference will be of $O(\varepsilon^2)$.
Mean PV evolves in shallow water due to the correlation between PV and divergence:
\begin{align}
    \pe_t \mean{q} = \mean{(u_x+v_y)q}. 
\end{align}
The weak time tendency implies that this correlation term is weak, mostly likely due to the fact that divergence is weak. In fact, for SWQG\pl~evolution, the mean of the advected PV ($\mean{q}$) does not change. Indeed,
\begin{subequations}
\begin{align}
    \mean{ (u_x+v_y) q } &= \mean{ (F^1_x+G^1_y)(\mcal{S}\Phi^0 + \mean{q}) }\\
    &= \mean{ \mcal{S}(F^1_x+G^1_y)\Phi^0 }\\
    &= \mean{ J(\nabla^2\Phi^0,\Phi^0)\Phi^0} = 0 .
\end{align}
\end{subequations}
Therefore, $C_q$ is the only representation of the evolution of the mean of PV ($\mean{q}$) in the SWQG\pl~model.

\section{Multi-layer shallow water}\label{sec:multi}
We extend the QG\pl~model to a multi-layer configuration. Here we only present the two-layer version, but the extension to more layers is straightforward, following the steps laid out here. A simulation of the two-layer version of SWQG\pl~under baroclinic instability shows that it can capture the vortex asymmetry in nonlinear baroclinic waves.

\subsection{The two-layer shallow water system and the derivation of the two-layer SWQG\pl~model}\label{sec:multi_QGp1_deri}
We start with the two-layer shallow water system on the {$\beta$-plane} with two equal layers of height $H$ and inter-layer reduced gravity $g'$
\begin{subequations}\label{eq:twolaySW}
\begin{align}
    &\Ro\frac{\DD_1 \ve u_1}{\DD t}+( f+\Bt \beta y)\ve{e}_z\times \ve u_1 = -g\nabla(h_1+h_2),\\
    &\Ro\frac{\DD_2 \ve u_2}{\DD t}+( f+\Bt \beta y)\ve{e}_z\times \ve u_2 = -g\nabla(h_1+h_2)-g'\nabla h_2,\\
    &\Ro\left[\frac{\DD_1 h_1}{\DD t}+h_1\nabla\cdot \ve u_1\right]+\Bu H\nabla\cdot \ve u_1 = 0,\\
    &\Ro\left[\frac{\DD_2 h_2}{\DD t}+h_2\nabla\cdot \ve u_2\right]+\Bu H\nabla\cdot \ve u_2 = 0.
\end{align}
\end{subequations}
where we scale the perturbation height geostrophically:
\begin{align}
    &h_1+h_2\sim\frac{fUL}{g}\\
    &h_1\sim h_2\sim\frac{fUL}{g'}
\end{align}
All nondimensional numbers are the same as those for the one-layer shallow water, except we use the baroclinic Burger number
\begin{align}
    Bu := \frac{g'H}{f^2L^2}.
\end{align}
The system conserves the shallow water PV in each layer ($i=1,2$):
\begin{subequations}\label{eq:mult_PV}
\begin{align}
    HQ_i =& \frac{f+\Ro (\zeta_i+\Gm \beta y)}{1+\{\varepsilon/\textit{Bu}\} h_i/H}\\
    =& f+\Ro\left[(\zeta_i+\Gm \beta y)-\Ub\frac{fh_i}{H}\right]\\
    &\qquad+\Ro^2\left[\Ub^2\frac{fh_i^2}{H^2}-\Ub\frac{(\zeta_i+\Gm \beta y)h_i}{H}\right]+O
    \left(\Ro^3\right)
\end{align}
\end{subequations}
where we expanded in the small Rossby number.

The potential form for the two-layer shallow water velocity follows from the one-layer version. The $\Phi$ components of the height are modified to the familiar QG form. 
\begin{subequations}\label{eq:stream_twolay}
\begin{align}
    u_1 &= -\Phi_{1,y}-F_1  \\
    u_2 &= -\Phi_{2,y}-F_2  \\
    v_1 &=  \Phi_{1,x}-G_1  \\
    v_2 &=  \Phi_{2,x}-G_2  \\
    h_1 &=  \frac{f}{g'}(\Phi_1-\Phi_2) + \frac{f}{g}\Phi_1-\Bu\frac{H}{f}G_{1,x}+\Bu\frac{H}{f}F_{1,y}  \\
    h_2 &=  \frac{f}{g'}(\Phi_2-\Phi_1)-\Bu\frac{H}{f}G_{2,x}+\Bu\frac{H}{f}F_{2,y}.
\end{align}
\end{subequations}

To derive SWQG and SWQG\pl, we expand the potential form \eqref{eq:stream_twolay}, similar to \eqref{eq:streamexp_onelay}.
The zeroth order of the PV expression \eqref{eq:mult_PV} gives the PV inversion of SWQG
\begin{align}
    \mcal{L}\begin{pmatrix}
        \Phi_1^0\vspace{3mm}\\
        \Phi_2^0
    \end{pmatrix} = \begin{pmatrix*}[l]
        \nabla^2\Phi_1^0+\Ub\frac{f^2}{g'H}(\Phi^0_2-\Phi^0_1)-\Ub\frac{f^2}{gH}\Phi_1^0\vspace{1mm}\\
        \nabla^2\Phi_2^0+\Ub\frac{f^2}{g'H}(\Phi^0_1-\Phi^0_2)
    \end{pmatrix*} = \begin{pmatrix*}[l]
        q_1\vspace{3mm}\\
        q_2
    \end{pmatrix*}-\Gm\beta y.
\end{align}
The operator on the left-hand side will appear in all higher-order inversions, and we name it $\mcal{L}$ for convenience.

We follow the same derivation steps as in Section \ref{sec:sing_QGp1_deri} but with more algebra manipulations. Inversions for $\Phi^1$'s come from the first order of the PV expression \eqref{eq:mult_PV}
\begin{align}
    \mcal{L}\begin{pmatrix}
        \Phi_1^1\vspace{3mm}\\
        \Phi_2^1
    \end{pmatrix} = \begin{pmatrix*}[l]
        C_{q,1}-\left[\Ub^2\frac{f}{H^2}\left(\frac{f}{g'}(\Phi^0_1-\Phi^0_2)+\frac{f}{g}\Phi^0_1\right)^2\right.\\
        \left.\qquad -\Ub\frac{1}{H}(\nabla^2\Phi^0_1+\Gm \beta y)\left(\frac{f}{g'}(\Phi^0_1-\Phi^0_2)+\frac{f}{g}\Phi^0_1\right)\right]
        \vspace{1mm}\\
        C_{q,2}-\left[\Ub^2\frac{f}{H^2}\left(\frac{f}{g'}(\Phi^0_2-\Phi^0_1)\right)^2\right.\\
        \left.\qquad -\Ub\frac{1}{H}(\nabla^2\Phi^0_2+\Gm \beta y)\left(\frac{f}{g'}(\Phi^0_2-\Phi^0_1)\right)\right]
    \end{pmatrix*}
\end{align}
The $C_{q,i}$'s are again to make sure $\mean{\Phi^1_i}=0$.

The elliptic problems for the rest of the potentials $F$ and $G$ come from appealing to the thermal wind balance. We form the so-called ``imbalance'' equation using the $O(\varepsilon)$ equations. Then one side becomes elliptic operators applied to the potentials, while the other side only depends on $\Phi^0_{1,2}$. We leave the details of the derivation to Appendix \ref{app:FG_deri}. The inversions 
for $F^1_1,F^1_2$ are
\begin{align}\label{eq:ellip_Fboth}
    \mcal{L}\begin{pmatrix}
        F^1_1\vspace{3mm}\\F^1_2
    \end{pmatrix} &= \Ub\frac{f}{H}\begin{pmatrix*}[l]
        \left.\left[J(\Phi^0_{1,x}+\Phi^0_{2,x},\Phi^0_{1}-\Phi^0_{2})+\Gm\beta y (\Phi_{1}^0-\Phi_{2}^0)_{,y}\right]\right/g'\\
        +\left[J(\Phi^0_{1,x},\Phi^0_{1})+\Gm\beta y \Phi_{1,y}^0\right]/g \vspace{2mm}\\
        \left.\left[J(\Phi_{1,x}^0+\Phi_{2,x}^0,\Phi^0_2-\Phi^0_1)+\Gm\beta y (\Phi_{2}^0-\Phi_{1}^0)_{,y}\right]\right/g'
    \end{pmatrix*},
\end{align}
and for $G_1^1,G_2^1$ are
\begin{align}\label{eq:ellip_Gboth}
    \mcal{L}\begin{pmatrix}
        G^1_1 \vspace{3mm}\\G^1_2
    \end{pmatrix} &= \Ub\frac{f}{H}\begin{pmatrix*}[l]
        \left[J(\Phi^0_{1,y}+\Phi^0_{2,y},\Phi^0_1-\Phi^0_2)+\Gm\beta y (\Phi_1^0-\Phi_2^0)_{,x}\right]/g'\\
        +\left[J(\Phi^0_{1,y},\Phi^0_1 )-\Gm\beta y\Phi_{1,x}^0\right]/g \vspace{2mm}\\
        \left[J(\Phi_{1,y}^0+\Phi_{2,y}^0,\Phi^0_2-\Phi^0_1)+\Gm\beta y(\Phi^0_2-\Phi^0_1)_{,x}\right]/g'
    \end{pmatrix*}.
\end{align}
The multi-layer SWQG\pl~inversions again have the same elliptic problems for the potentials.

\subsection{Baroclinic unstable jets simulations}
To explore the modeling power of multi-layer SWQG\pl, we simulate the evolution of a baroclinically unstable jet. Baroclinic instability is only possible in shallow water with at least two layers. It spins up vorticity features from small noise, at the expense of large-scale potential energy. This is in contrast and in complement to the freely-decaying set-up in Section \ref{sec:free_sim}.

\subsubsection{Set-ups of the simulations}
We set our simulation in the atmospheric context, inspired by the work of \citet{LambaertsEtAl_12}. The idealized two layers atmosphere has $f=10^4\text{ s}^{-1}$, $g'=2\text{ m/s}^2$, $H=10$ km {mean thickness}, and typical flow speed of $U=5\text{ m/s}$. If we nondimensionalize with a length scale of $L=500$ km, the nondimensional number will be $Bu = 8$ and $\varepsilon=0.1$. The domain is doubly-periodic with $Ly=32\times L$ and $Lx = 64\times L$. 

\begin{figure}
    \centering
    \includegraphics{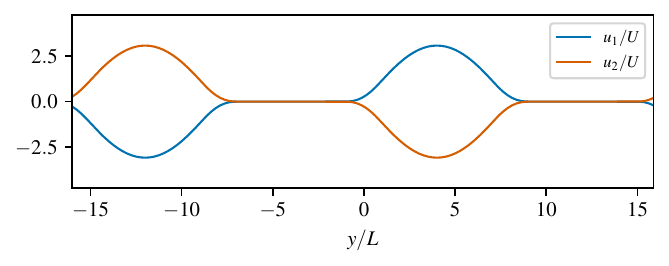}
    \caption{The initial jets' nondimensional velocity in the upper ($u_1/U$) and lower ($u_2/U$) layer.}
    \label{fig:init_bickely_jet}
\end{figure}

Initially, the domain is filled with two large baroclinic jets of meridional size $4000$ km ($8\times L$) of maximum speed 13 m/s, as shown by Figure \ref{fig:init_bickely_jet}. The northern one has a westerly jet in the top layer and an easterly jet in the bottom layer, a rough sketch of the polar jet stream. The southern jet is the opposite and is present only to restore periodicity for the height perturbation. We will focus on the northern jet. The initial heights are set to match the initial jets in geostrophic balance. Since this initial condition only varies in one direction, it is already a solution of the two-layer shallow water model and in perfect balance. No divergence field correction is needed. 

The instability is triggered by the seeding of small magnitude noise, to height in the shallow water simulation, and to PV in the SWQG\pl~simulation. We again use Dedalus to simulate the nonlinear evolution. The resolution is $Nx\times Ny = 512\times 256$ modes. Everything else about the numerics are the same as the simulation in Section \ref{sec:free_sim} except the fourth-order hyper-dissipation constant is larger {with non-dimensional value}
\begin{align}
    \nu=7.3\times 10^{-4}. 
\end{align}

\subsubsection{Simulation results}
\begin{figure}
    \centering
    \includegraphics{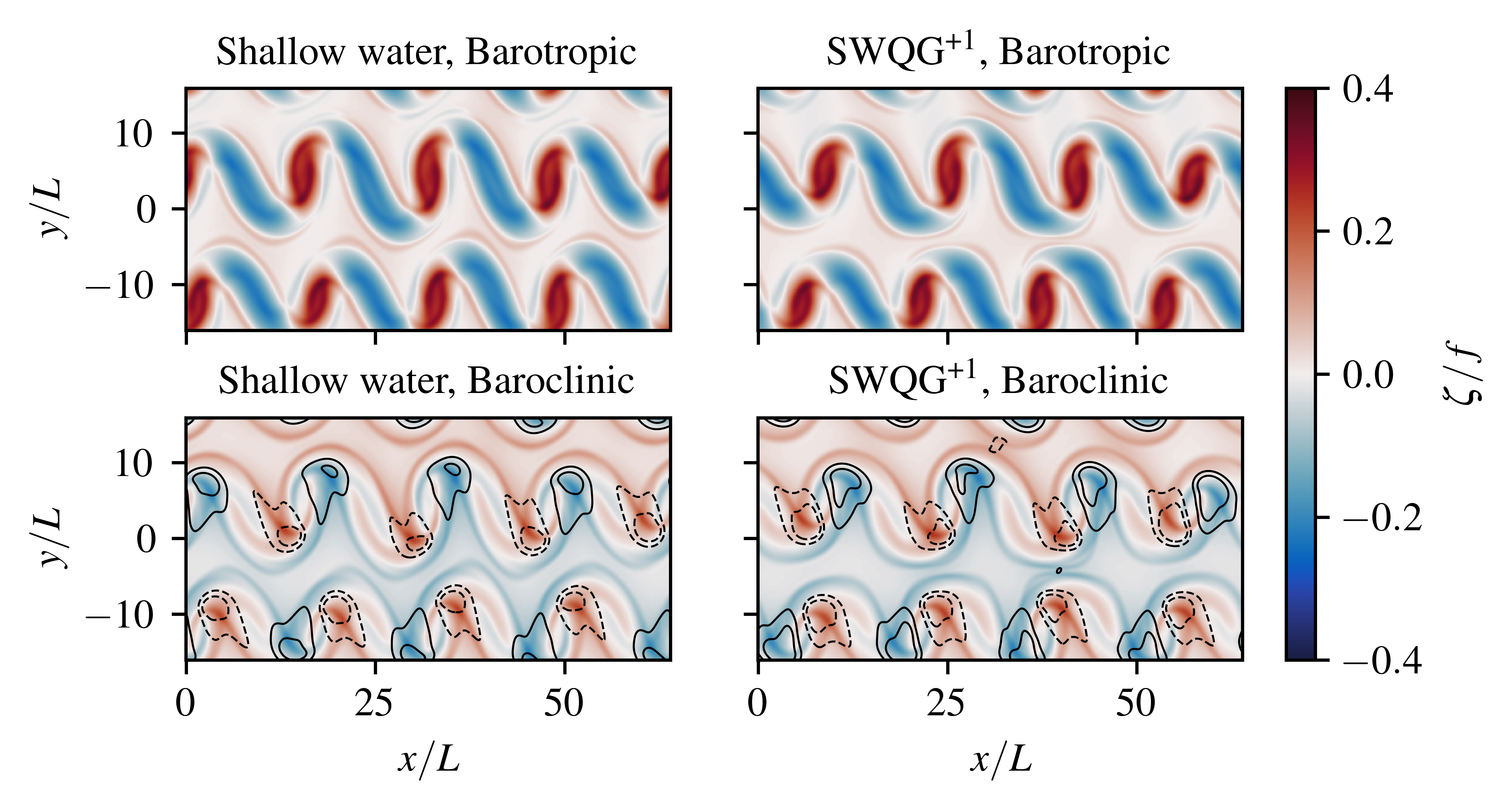}
    \caption{The vorticity field of the evolution of the jets for the shallow water model (left) and SWQG\pl~model (right). The contour is divergence $\Ro\delta/f$ at $[-0.1,-0.005,0.005,0.1]$, with the solid being positive. The shallow water snapshot is taken at $t/T=30$ while the SWQG\pl~snapshot is taken at $t/T=24$. }
    \label{fig:twolay_zeta_snap_t27d0}
\end{figure}

\begin{figure}
    \centering
    \includegraphics{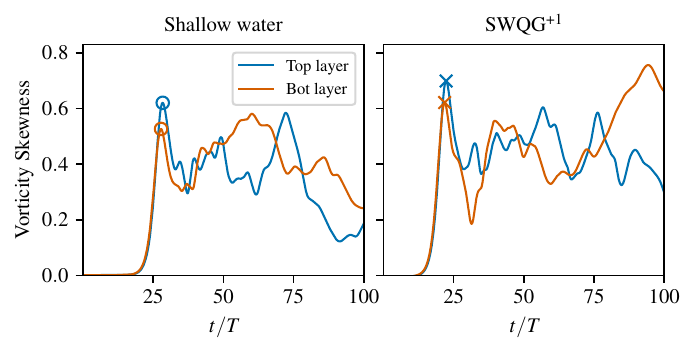}
    \caption{Vorticity skewness of two-layer shallow water (left) and SWQG\pl~(right) during the evolution of unstable jets simulations. {The circles and crosses mark the first local maxima of vorticity skewness, which is further explored in Figure \ref{fig:MaxSkew_Rosweep}.}}
    \label{fig:two_lay_vortskew}
\end{figure}

The SWQG\pl~captures the vorticity evolution of the baroclinic life cycle to remarkable accuracy. Figure \ref{fig:twolay_zeta_snap_t27d0} displays snapshots of the barotropic ($(\zeta_1+\zeta_2)/2$) and baroclinic ($(\zeta_1-\zeta_2)/2$) vorticity (in color) and divergence (in contour) during the growth of the baroclinic instability. Note that the time of the snapshot is offset. This is because the growth of the instability from small-scale noise happens at different times in the model, due to the chaotic nature of the instability. {However, the growth rate is remarkably similar (see Figure \ref{fig:two_lay_vortskew}).} The match is indeed impressive. Not only does the SWQG\pl~model capture the asymmetric evolution of the cyclonic and anticyclonic part of the baroclinic wave, it is also able to recover the associated divergence fields. {Note that one could infer the same divergence, using the $\omega$-equation. The difference of SWQG\pl~is that the full velocity participates in the advection, while in SWQG, the higher-order divergence is implicit and is not used in the advection of PV.}
The divergence is of crucial atmospheric interest as a convergent region in the lower layer corresponds to precipitation. For our example, baroclinic convergence correlates with baroclinic cyclonic vorticity, and baroclinic divergence is large near regions of barotropic strain. On the other hand, barotropic divergence is uniformly small. This is reminiscent of the strain-induced fronts configuration studied in \citet{HoskinsBretherton_72}. We will explore further strain-induced fronts in Section \ref{sec:front}. {The similarity of the evolution for the two models go beyond this early snapshot. In the supplementary materials, we provide videos of the $\Ro=0.1$ simulations.}

\begin{figure}
    \centering
    \includegraphics{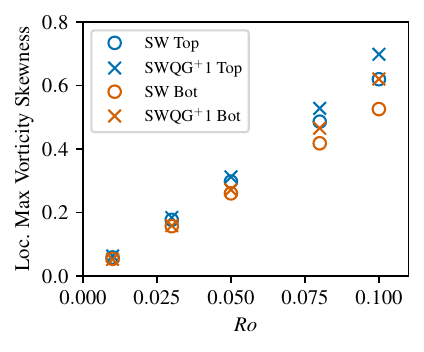}
    \caption{{The first local maxima of vorticity skewness of a set of simulations with varying Rossby numbers.}}
    \label{fig:MaxSkew_Rosweep}
\end{figure}

The match of the vorticity snapshots translates to the ability of the SWQG\pl~to capture the vorticity skewness of the simulation. Figure \ref{fig:two_lay_vortskew} shows the agreement of the overall trend {over} a large time ($t/T=100$), for both layers. The vorticity is now skewed towards positive, in comparison to the freely decaying result. This change in vorticity skewness is due to the new effects of the baroclinic mode now included in the two-layer model. The vortex stretching due to the correlation of baroclinic vorticity and divergence drives the initial growth of positive vorticity skewness shown in Figure \ref{fig:two_lay_vortskew}. This effect is entirely missing in the one-layer simulation. {The vorticity skewness is an ageostrophic phenomenon. We diagnose the first local maxima of the vorticity skewness, as it is an indicator of the strength of the ageostrophic frontogenesis. Figure \ref{fig:MaxSkew_Rosweep} shows clearly that this measurement of vorticity skewness is linearly proportional to the Rossby number, and SWQG\pl~is effective over a large parameter space.} 

Our results agree with those of \cite{LambaertsEtAl_12} in the upper layer for the onset of instability. However, their lower layer does not have a mean jet while ours does, thus the results differ. 

\subsection{Baroclinic strain-driven fronts modeled by SWQG\pl}\label{sec:front}
The two-layer simulation reveals that strain-driven fronts are important in the dynamics and contribute to the positive vorticity skewness{, which is opposite to the results of our one-layer simulations}. {This is a new feature of the two-layer shallow water simulations and worth further investigation.} Here we study a minimal model of strain-driven fronts by adapting the set-up of \cite{HoskinsBretherton_72} to the two-layer model. 

We study a strain{-driven} cold filament across the $y$-direction. That is, we ignore the $x$-variation for all perturbation fields. The filament has a center with a large $h_2$ perturbation. We prescribe the perturbation PV field to be Gaussian
\begin{align}
    q_1(y) = -q_2(y) = e^{-y^2}
\end{align}
and solve for the physical fields under the response of a barotropic incompressible, irrotational strain field
\begin{align}
    {U}^M=\alpha x, \qquad {V}^M = -\alpha y,
\end{align}
which can be captured by a horizontal streamfunction
\begin{align}
    {\Phi}^M = -\alpha xy.
\end{align}

The strain field has no imprint on the PV if we take the rigid-lid limit of SWQG\pl. That is, we ignore all terms that are divided by $g$. All inversions are not affected by the strain field except for $G$ (and $F=0$), where we have

\begin{align}
    \mcal{L}\begin{pmatrix}
        G^1_1 \vspace{3mm}\\G^1_2
    \end{pmatrix} &= \Ub\frac{2\alpha f}{g'H}\begin{pmatrix*}[l]
        (\Phi^0_2-\Phi^0_1)_{,y} \vspace{2mm}\\
        (\Phi^0_1-\Phi^0_2)_{,y}
    \end{pmatrix*}.
\end{align}
Note that the operator $\mcal{L}$ now only has $y$-derivatives. Without the strain field ($\alpha=0$), there would be no response since $G=0$. This is expected since the filament is a geostrophically balanced solution to the shallow water equations. 

The \citet{HoskinsBretherton_72} model is posed in an infinite domain in $y$, where perturbation velocities decay at infinity. This implies boundary conditions for the potentials:
\begin{align}
    &\left.G^{1}\right|_{y=\pm\infty} = 0\\
    &\left.\Phi^{0,1}_{,y}\right|_{y=\pm\infty} = 0.
\end{align}
Numerically, we solve for the potentials in a finite domain $y/L\in[-5,5]$, which is large enough so that the solution is not changed with a large domain. It is represented numerically using $N=256$ Chebyshev modes, using the Dedalus solver. 

\begin{figure}
    \centering
    \includegraphics{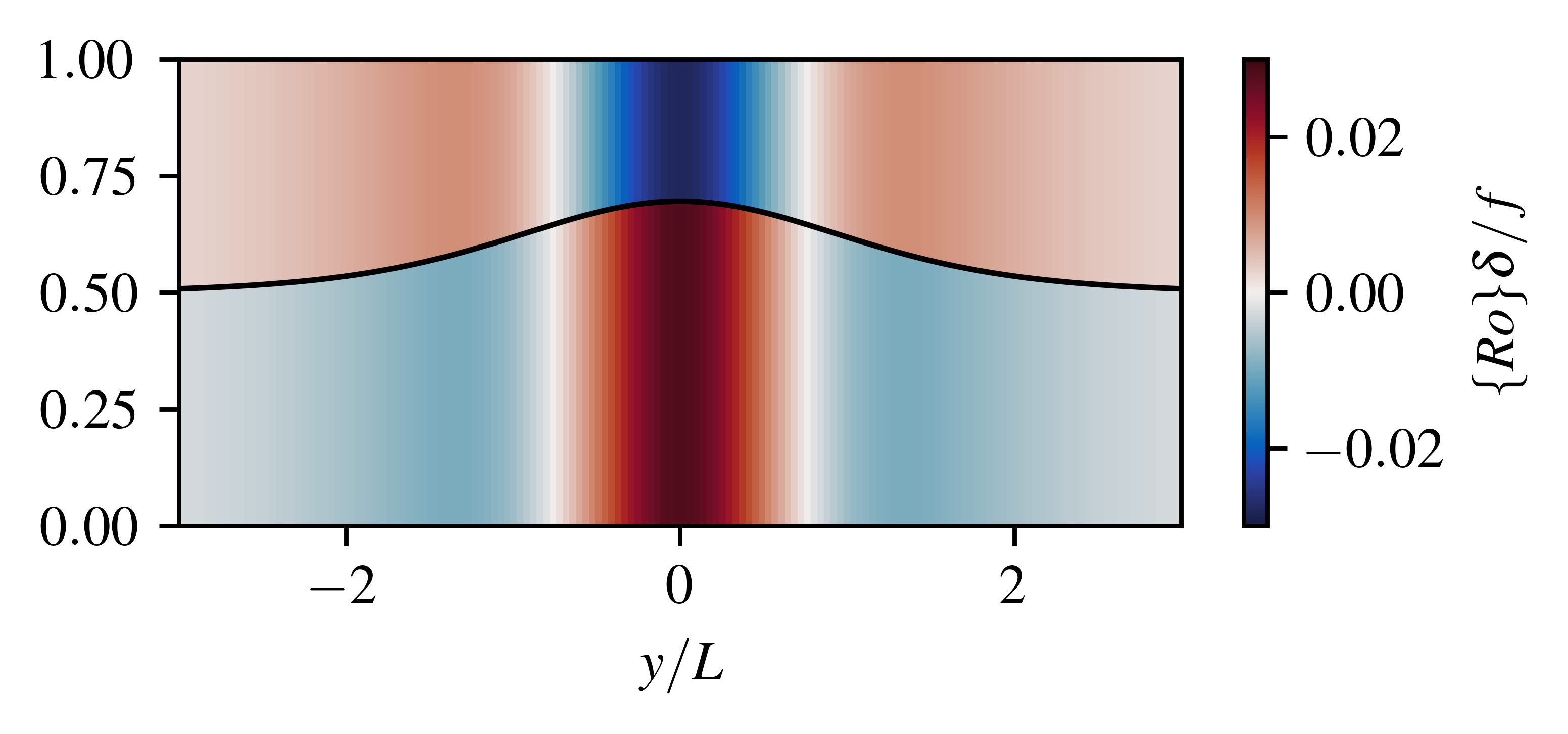}
    \caption{Divergence of a cold filament driven by a barotropic strain, modeled by a two-layer rigid lid SWQG\pl~model. The domain is divided vertically into two layers by the height perturbation. The color shows divergence values. This is a zoomed-in view showing $y/L = [-3,3]$}
    \label{fig:front_div}
\end{figure}

We solve the system with $Bu=1$ and $\varepsilon=0.3$. Figure \ref{fig:front_div} shows the divergence
\begin{align}
    \delta_1 = v_{1,y} &=  -\Ro G_{1,y} \qdt{and} \delta_2 = v_{2,y} =  -\Ro G_{2,y}.
\end{align}
overlaid on the height perturbations. In the top layer, a cold filament has cyclonic vorticity, which is correlated with convergence. This correlation is a key feature of fronts and explains the cyclonic bias of vorticity in the two-layer simulations by appealing to vortex stretching{: baroclinic cyclonic vorticity is correlated with baroclinic negative divergence}. {Note that the above equation is equivalent to the ``$\omega$-equation''. Therefore, it is implicit in the SWQG model. However, SWQG does not use this ageostrophic velocity in its advective evolution. It does not capture the vorticity skewness evolution.}

{Finally, a comparison with the surface buoyancy gradient driven Boussinesq frontogenesis of \cite{HoskinsBretherton_72} is warranted. }
In the layered shallow water set-up, the convergence will smooth away the height perturbation and release the stored APE. Shallow water fronts would be less strong with no finite time singularity, compared to the Boussinesq case. 

\section{Conclusion and outlook}\label{sec:conclude}
{In this paper, we have recapitulated and derived the PV-based balanced model, SWQG\pl, for one and multiple-layer shallow water systems. Through simulations, we have shown that SWQG\pl~}can capture the balanced ageostrophic effects of the shallow water model in the finite Rossby number regime. In particular, it can model {the negative skewness of vorticity in freely-decaying one-layer shallow water, as well as the positive skewness in the baroclinic instability of two-layer shallow water. It does this by} capturing the correlation between vorticity, divergence, and height for coherent vortices and baroclinic fronts. The {time evolution of the} energy and potential enstrophy {is close to that of the full shallow water model}, even though we cannot prove exact conservation. This shows that SWQG\pl~{has potential as a tool for the} study of many GFD problems in the atmosphere and ocean on Earth and other planets. {Its dynamical evolution is free of inertial-gravity waves by design, focusing the model on transport active balanced motion. The diagnostic relations between ageostrophic physics can be useful in inverse problems when observed physics is limited.}
There is much work to be done to adapt and extend the SWQG\pl~model. The rest of this section describes some possible applications and the desirable extensions to the model.

One central challenge in the parametrization of ocean eddies is to model correctly the mean location and variability of the ocean jets \citep{ChassignetMarshall_08}. \citet{ThiryEtAl_24} show that SWQG is inaccurate in modeling the jet location of a barotropic wind-driven gyre, perhaps the simplest model of the general ocean circulation. Therefore, QG does not capture enough phenomenology to be sufficient for eddy parameterization. However, PV-based thinking and balanced thinking have enabled much progress in eddy parametrization and should not be abandoned. SWQG\pl~offers a nice intermediate option. It is still simple like SWQG where one prognostic PV variable governs the entire model evolution. However, it is able to capture the important balanced ageostrophic effects. Future work should aim to replicate the asymmetric lean of the jet in a wind-driven gyre in the SWQG\pl~model. Boundary conditions for the SWQG\pl~model can be adapted from the boundary condition specifications in \citet{MurakiEtAl_99}.

The shallow water model {in one-layer \citep{LahayeZeitlin_16} and two-layer \citep{LambaertsEtAl_12,BembenekEtAl_20}} has been a useful simple model to study the nonlinear and non-smooth effects of moisture in the {mid-latitude} atmosphere. SWQG has also been adapted to include moisture effects. In particular, definitions of moist PV have been useful to interpret model results \citep{LapeyreHeld_04,SmithStechmann_17,LutskoHell_21,BrownEtAl_23}. SWQG\pl, with an appropriate definition of moist PV, can be a useful bridge between these two models. The removal of gravity waves removes an undesirable complication in the study of the large-scale evolution of storm tracks. More broadly, it would be interesting to see if moist effects can be sufficiently captured by inverting from the moist PV. That is, are moist effects fundamentally ``balanced'' or not?

The research on the effect of ocean topography on the coherent structure of the ocean eddy has enjoyed a resurgence \citep{SolodochEtAl_21,SiegelmanYoung_23,LaCasceEtAl_24}. Theoretical understanding of this problem typically comes from SWQG, or the even simpler two-dimensional Euler equation \citep[e.g.][]{BrethertonHaidvogel_76,SiegelmanYoung_23}. However, the necessary symmetry between different signed vortices in the SWQG model is unrealistic. The study of the realistic formation and destruction of coherent vortices requires numerical simulation using complex ocean models. SWQG\pl~and Boussinesq QG\pl~can bridge this gap. SWQG\pl~{improves upon} SWQG and allows for the {modeling of} realistic evolution of the coherent vortices {(though it still assumes asymptotically small topography height)}. In the meantime, there is still only one variable to predict, PV, and it is possible that many PV-based theoretical thinking can be applied to SWQG\pl. Of course, the study of coherent structure on other planets can also benefit from upgrading to the SWQG\pl~model \citep{SiegelmanEtAl_22}. For these applications, SWQG\pl~model needs to be extended and tested on the cases with non-constant $f$, topography, {and on spherical domains \citep{ChoPolvani_96}}.

From these examples, it is clear that there are many applications of SWQG\pl~and much work to be done. The results in this paper serve as the bare minimum showcase of its ability to capture phenomena of interest of one fundamental model of GFD, the shallow water model.


\backsection[Acknowledgements]{We thank Oliver B\"uhler, Guillaume Lapeyre, and Olivier Pauluis for helpful discussions, without implying their endorsement. This work was supported in part through the NYU IT High Performance Computing resources, services, and staff expertise. We thank Shenglong Wang for his work on configuring Dedalus on the NYU Greene cluster. We thank three anonymous reviewers for their helpful suggestions. }

\backsection[Funding]{KSS and RD acknowledge support from NASA contract 80NSSC20K1142.}

\backsection[Declaration of interests]{The authors report no conflict of interest.}

\backsection[Data availability statement]{The code that generates the data and figures in this paper is available
on GitHub: \url{https://github.com/Empyreal092/SWQGp1_freedecay_BCIjets_Public}.}

\backsection[Author ORCIDs]{R. S. D\`u, https://orcid.org/0000-0001-8201-1851}

\backsection[Author contributions]{}

\appendix

\section{Equivalency of the SWQG\pl~potential form to the model of V96}\label{app:V96}
We show that our model is equivalent to the model in \citet{WarnEtAl_95} and the extended version under the $\beta$-approximation in \S 2(a) of V96 up to the constant terms in the PV inversions. 

First, the two models are the same at the QG level. Then the V96 model inverts for the next-order quantities using their {(V15-V17)}. We show that the two models are equivalent by recovering {(V15-V17)} from the SWQG\pl~inversions. V96's {(V15)} is the next-order PV inversion by noting its left-hand-side is $\mcal{S}(\Phi^1)$ while the right-hand-side is equal to the right-hand-side of \eqref{eq:SWpl_Phi1inv}. {For (V17), the shallow water $\omega$-equation, we have
\begin{align}
    \mcal{S}(u_x+v_y) &= -\mcal{S}(F^1_x+G^1_y)\\
    &= \Ub\frac{f}{gH}\left[J(\Phi^0,\nabla^2\Phi^0)+\Gm \beta \Phi_x^0\right].
\end{align}
}

{To recover (V16), the Bolin–Charney balance equation \citep{Bolin_55,Charney_55a}, we have
\begin{align}
    f \zeta^1-g\nabla^2 \eta^1 &= f (\nabla^2\Phi^1-G^1_x+F^1_y) - f\nabla^2\Phi^1 - \Bu\frac{gH}{f}(-\nabla^2 G^1_x+\nabla^2 F^1_y)\\
    &= \Bu\frac{gH}{f}\left[\mcal{S}(G^1_x)-\mcal{S}(F^1_y)\right]\\
    &= -2J(\Phi^0_x,\Phi^0_y)-\Gm \beta \Phi_y^0-\Gm \beta y \nabla^2\Phi^0
\end{align}
where we used the elliptic problem for $F^1$ and $G^1$. The last term on the right-hand side is not included in (V16). This is a mistake in V96 which stems from the same term missing in the ``divergence equation'' (V6). Forming the divergence of the shallow water momentum equations \eqref{eq:oneSW} reveals (V6) should additionally include a $-\beta y \zeta$ term.
}

\section{The derivations for the elliptic inversions of $F_1^1$, $F_2^1$, $G_1^1$, and $G_2^1$ for the two-layer QG\pl~model}\label{app:FG_deri}

{We start with the two-layer shallow water system \eqref{eq:twolaySW} and the potential form for the physical variables in the two-layer case \eqref{eq:stream_twolay}. For the QG-level variables, they are in thermal wind balance. For $h_{1,x}$, this is}
\begin{align}
    h^0_{1,x} &= \frac{f}{g'}(\Phi^0_{1,x}-\Phi^0_{2,x})+\frac{f}{g}\Phi^0_{1,x}= \frac{f}{g'}(v^0_1-v^0_2)+\frac{f}{g}v^0_1.
\end{align}
The ``imbalance'' equation {is the time evolution equation of the difference between the thermally balanced variables, that is, the ``imbalance''. For the QG-level geostrophically balanced variables, the difference should be zero for all time. This allows us to form a diagnostic equation free of time partial derivatives.}
\begin{align}\label{eq:imba_F1}
    &\left(\frac{\DD^0_1 h^0_1}{\DD t}\right)_x - \frac{f}{g'}\left(\frac{\DD^0_1 v_1^0}{\DD t}-\frac{\DD^0_2 v_2^0}{\DD t}+\Gm\beta y u_1^0-\Gm\beta y u_2^0\right)-\frac{f}{g}\left(\frac{\DD^0_1 v_1^0}{\DD t}+\Gm\beta y u_1^0 \right)\nonumber\\
    =& -\Bu H(u^1_{1,xx}+v^1_{1,xy})- \frac{f}{g'}\left( -fu_1^1+fu_2^1 \right)-\frac{f}{g}\left( -fu_1^1-gh^1_{1,y} \right).\\
    \Rightarrow \quad  &\nabla^2 F^1_{1}+ \Ub\frac{f^2}{g'H}\left( F_2^1-F_1^1 \right)-\Ub\frac{f^2}{gH} F_1^1 \\
    =& \Ub\frac{f}{H}\left(\left.\left[J(\Phi^0_{1,x}+\Phi^0_{2,x},\Phi^0_{1}-\Phi^0_{2})+\Gm\beta y (\Phi_{1}^0-\Phi_{2}^0)_{,y}\right]\right/g'\right.\\
    &\qquad\qquad\quad \left.\left.+\left[J(\Phi^0_{1,x},\Phi^0_{1})+\Gm\beta y \Phi_{1,y}^0\right]\right/g\right).\label{eq:ellip_F1}
\end{align}
We see that the right-hand side becomes elliptic operators applied to the potentials, while the left-hand side only depends on $\Phi^0_{1,2}$.  

For $F_2^1$, we form {the ``imbalance'' equation}
\begin{align}
    &\left(\frac{\DD^0_2 h^0_2}{\DD t}\right)_x - \frac{f}{g'}\left(\frac{\DD_2^0 v_2^0}{\DD t}-\frac{\DD_1^0 v_1^0}{\DD t}+\Gm\beta y u_2^0-\Gm\beta y u_1^0\right) \\
    =& -\Bu H(u^1_{2,xx}+v^1_{2,xy})- \frac{f}{g'}\left( -fu_2^1-g'h^1_{2,y}+fu_1^1 \right).
\end{align}
{Similar steps gives}
\begin{align}
    &\nabla^2 F^1_{2}+ \Ub\frac{f^2}{g'H}\left( F_1^1-F_2^1 \right) \\
    =& \Ub\frac{f}{g'H}\left[J\left(\Phi_{1,x}^0+\Phi_{2,x}^0,\Phi^0_2-\Phi^0_1\right)+\Gm\beta y (\Phi_{2}^0-\Phi_{1}^0)_{,y}\right]
\end{align}
Together with \eqref{eq:ellip_F1} gives the compact form \eqref{eq:ellip_Fboth}.

For $G_1^1$ and $G_2^1$, 
\begin{align}
    &\left(\frac{\DD^0_1 h^0_1}{\DD t}\right)_y - \frac{f}{g'}\left(\frac{\DD^0_2 u_2^0}{\DD t}-\frac{\DD^0_1 u_1^0}{\DD t}-\Gm\beta yv_2^0+\Gm\beta yv_1^0\right)+\frac{f}{g}\left(\frac{\DD^0_1 u_1^0}{\DD t}-\Gm\beta yv_1^0 \right)\\
    =& -\Bu H(u^1_{1,xy}+v^1_{1,yy})- \frac{f}{g'}\left( fv_2^1-fv_1^1 \right)+\frac{f}{g}\left(fv_1^1-gh^1_{1,x}\right)\\
    &\left(\frac{\DD^0_2 h^0_2}{\DD t}\right)_y - \frac{f}{g'}\left(\frac{\DD^0_1 u_1^0}{\DD t}-\frac{\DD^0_2 u_2^0}{\DD t}-\Gm\beta y v_1^0+\Gm\beta y v_2^0\right) \\
    =& -\Bu H(u^1_{2,xy}+v^1_{2,yy})- \frac{f}{g'}\left( fv_1^1-fv_2^1+g'h^1_{2,x} \right).
\end{align}
Standard but somewhat tedious algebraic manipulation, using the shallow water version of the thermal wind balance gives \eqref{eq:ellip_Gboth}.

\bibliographystyle{jfm}
\bibliography{jfm}

\end{document}